# Generic and adaptive probabilistic safety assessment models: Precursor analysis and multi-purpose utilization

Ali Ayoub[1,a,b], Wolfgang Kröger[a], and Didier Sornette[a,b,c]

[a] Chair of Entrepreneurial Risks, D-MTEC, ETH Zürich, Switzerland
[b] ETH Risk Center, ETH Zürich, Switzerland
[c] Institute of Risk Analysis, Prediction and Management (Risks-X), Academy for Advanced Interdisciplinary Studies, Southern University of Science and Technology (SUSTech), Shenzhen, 518055, China

ABSTRACT

Motivated by learning from experience and exploiting existing knowledge in civil nuclear operations, we have developed in-house generic PSA models for pressurized and boiling water reactors. The models are computationally light, handy, transparent, user-friendly, and easily adaptable to account for major plant-specific differences. They cover the common internal initiating events, frontline and support systems reliability and dependencies, human-factors, common-cause failures, and account for new factors typically overlooked in many developed PSAs. For quantification, the models use generic US reliability data, precursor analysis reports and studies, the ETHZ Curated Nuclear Events Database, and experts' opinions. Moreover, uncertainties in the most influential basic events are addressed. The generated results show good agreement with assessments available in the literature with detailed PSAs. We envision the models as an unbiased framework to measure nuclear operational risk with the same "ruler", and hence support inter-plant risk comparisons that are usually not possible due to differences in plant-specific PSA assumptions and scopes. The models can be used for initial risk screening, order-of-magnitude precursor analysis, and other research and pedagogic applications especially when no plant-specific PSAs are available. Finally, we are using the generic models for large-scale precursor analysis that will generate big picture trends, lessons, and insights.

## 1. Introduction

Due to the rare nature of severe accidents in the nuclear industry (core damage and large releases), it is very hard to quantify its risk on a pure empirical basis. Therefore, probabilistic frameworks such as Probabilistic Safety Assessment (PSA) [1] have provided a sound alternative for risk estimation by system decomposition and making use of failure data at a more basic level (e.g. components level) to derive system-level behavior. PSA is performed at three sequential levels [2]:

1- PSA Level 1 models the plant's response to initiating/perturbing events and aims at quantifying the risk of a core damage, namely, core damage frequency (CDF) per reactor-year.

---

[1] Corresponding author:
Email: aayoub@ethz.ch (A. Ayoub)
Phone: +41 44 632 24 45

2- PSA Level 2 models the respective containment response to the accident/initiator, and aims at quantifying the radioactive release to the environment, namely, large early release frequency (LERF) per reactor-year.
3- PSA Level 3 models the offsite consequences of the release, and aims at quantifying the risk to the public (frequency and impact on public health and the environment, as well as direct costs).

Moreover, PSA frameworks are used to quantify the risk of operational experience through precursor analysis. A precursor is an observed event resembling a truncated accident sequence (accident sub-chain) that could lead to an accident (e.g. core damage) if combined with additional adverse conditions [3]. Precursor analysis is a hybrid empirical-probabilistic method to estimate operational risks such as CDF, by mapping the observed event sequence to the PSA model, and calculating the remaining distance (probability) to core damage. Precursor analysis have been used by regulators around the world to monitor plants' risk over time [4], and provide an alternative robust estimate of CDF other than PSA CDFs [5].

Over the time, and through its extensive use – both for regulatory and risk-informed decision-making purposes – PSA witnessed rapid developments and reached high levels of details and sophistication. It evolved as a very plant-specific and site-specific method, capturing the very details of each plant. However, it remains difficult to use or understand outside the circle of their developers or super-experts [6]. The resulting complexities, in addition to the absence of a standardized PSA methodology and scope, made it difficult to compare results of different PSAs [7], and hindered possibilities for design-to-design and plant-to-plant safety comparisons. Therefore, PSAs became less suitable to understand industry-wide performance and big picture safety insights and trends.

Some researchers and organizations started efforts to establish generic PSA models or PSA platforms to facilitate exchanges and understanding. For example, the Open PSA Initiative [8, 9] started an open platform to encourage peer review and transfer of ideas and PSA lessons, and to reach a common PSA representation worldwide. The Spanish Nuclear regulator (CSN) has initiated a PSA harmonization activity to construct generic and standardized PSA models for the Spanish nuclear fleet to achieve a licensee-independent risk-view, targeted inspection activities, and unified platform for precursor analysis [10]. The Nordic PSA Group (NPSAG) started a project to harmonize PSA and improve its transparency and comparability [11, 12]. At EDF (Electricité de France), efforts are done to simplify PSA representation through modularization and "object-oriented" modeling [13].

In this work, we offer a step forward towards PSA harmonization and handiness. We present our final in-house PSA models for light water reactors (LWRs) initially introduced in [6, 14]. We have developed an open and generic PSA models for pressurized and boiling water reactors (PWRs and BWRs) in SAPHIRE (a probabilistic risk and reliability assessment software tool, which stands for *Systems Analysis Programs for Hands-on Integrated Reliability Evaluations*) [15]. The models cover all common internal initiating events, with intermediate complexity generic event trees and fault trees capturing important design differences without going to the detailed component level connections and layouts. This mentality of going simple and generic allowed for accounting for important contributors and lessons learnt from analyzing hundreds of precursors in our ETHZ nuclear events database [16, 17].

The developed PSA models proved to be well suited to perform efficient precursor analyses, providing representative order-of-magnitude risk estimates of operational events. The models will offer an unbiased framework to compare operational events and precursors at different plants by annealing-out many plant-specific differences as a result of going generic, hence forming a basis for inter-plants comparisons. By pooling worldwide nuclear operational experience, the models will allow us to understand big picture safety insights and trends. Our PSAs can be used to support plant-specific PSAs, and with their neat and simplified representation, they will provide an open and transparent framework that is ready to be employed when no access to detailed PSAs is provided (e.g. by research institutes, universities, NGOs and other organizations).

The organization of the manuscript is as follows. In Section 2, we present our modeling philosophy and approach. Sections 3 and 4 presents our detailed PWR and BWR models respectively. Section 5 discusses the quantification of the models. Section 6 presents applications of hands-on precursor analysis and examples. Section 7 concludes. An appendix at the end shows some modeled event trees and fault trees.

## 2. Modeling philosophy

Although LWRs differ in design and layout between different countries and vendors, nevertheless, the physics is basically the same and all plants utilize the same safety functions. In this work, we have developed generic, yet adaptive PSA models for both PWRs and BWRs, where each of the two models contains a set of linked event trees and fault trees of intermediate details. Each model is organized in a way suitable to describe all reactors of the same technology, with room for customization/adaptation to account for differences and design-specificities in available systems, configurations, degree of redundancy, human interventions and automation level, and support dependence. Depending whether a specific plant design has a specific safety system/function or not, its top events will be put to its nominal fault tree failure value or to a fail-state (Fail True). For example, the high pressure sump recirculation (HPR) function in PWRs is only available in some plants (mainly in US plants), and similarly for BWRs, with the differences in the available systems between the various construction lines such as the isolation condenser (IC), high pressure core spray (HPCS), and reactor core isolation cooling (RCIC).

Our nuclear events database [16] has been a major asset supporting many of our modeling needs, assumptions, and decisions. The developed models cover internal initiating events that are either very frequent, very serious, classical design-basis events, or appeared as precursors in our database. Furthermore, PSA level 1 event trees are developed for each initiating event, with two end states, 1- core damage (CD) when no sufficient core cooling is provided, and 2- success (OK) when the safety functions manage to cool the core and bring the reactor to a stable and safe state (within 24 hours).

Events and functions of the event trees are quantified with devoted fault trees that quantify failure probabilities at an aggregated train level -- without going into very components details. Support systems contributions are captured within the fault trees, and have both a global and a local element. Global support failures are failures on inter-system shared support trains, hence affecting trains in different safety systems (e.g. a shared AC bus), while local support failures are failures limited to a single train within a specific safety system (single pump component cooling circuitry or local activation/control logic circuity). Moreover, the fault trees include important human factors and human-related actions realized in our database [17], such as operator errors of omission and commission (EOO and EOC respectively), testing and maintenance (T&M) errors, and T&M unavailabilities. Common-cause failures (CCFs) are considered at two levels, a plant-level CCF[2] inferred from events within our database [18] that is modeled at the event tree level, and a typical system-level CCF modeled at the fault tree level. The conservative Beta-factor model was adopted for CCF quantification (details in section 5).

---

[2] Failures or potential failure in different systems, due to the same cause or some interactions, and this is typically not taken into account in PSA modeling. It includes deficits in safety culture, organizations, designs, and procedures, that tend to affect multiple systems, and sometimes the whole plant.

Recovery potentials are explicitly modeled in the fault trees as basic events, however, conservatively put to a failure state for most of the safety systems. Recovery of emergency diesel generators (EDGs) and some other support systems (e.g. offsite power) is an exception, as they are outside the containment and are generally, and realistically, easier to recover. Nevertheless, to be fair in our precursor analysis, cross-tying and recovery possibilities are credited whenever they are explicitly mentioned in the event following the logic used in the USNRC Accident Sequence Precursor program (ASP) [19].

Our models are limited to PSA level 1, hence generating CDF estimates only. For the LERF assessment (PSA level 2), we provide a rough estimate using the one-tenth rule of thumb, i.e. a conditional containment failure probability of 10% [20]; hence LERF is estimated as one-tenth of the CDF. This assumption was also used in the precursor analysis calculations, unless there was a containment bypass or a containment function failure during that precursor, in such a case, the LERF is equal to the CDF. Finally, our models are limited to the "at power" operation mode only.

**3. PWR models**

For PWRs, we have considered the following internal initiating events:

- General transients (including loss of main feedwater).
- Complicated transients:
    - Steam generator tube rupture (SGTR).
    - Loss of condenser heat sink (LOCHS).
    - Main Steam line break (MSLB).
- Loss of coolant accidents (LOCAs):
    - Small break LOCA (SBLOCA).
    - Medium break LOCA (MBLOCA).
    - Large break LOCA (LBLOCA).
- Total loss of support functions:
    - Loss of offsite power (LOOP).
    - Total loss of service water – both essential and normal (TLOSW).
    - Loss of normal service water (LONSW).

For each of these initiators, an event tree is developed, encompassing safety systems/functions and their backups, needed to mitigate core damage. Table 1 lists these generic safety functions and the respective modeled safety systems/top events.

Table 1. Generic PWR safety functions and the associated modeled safety systems/top events

| Safety Functions | Safety Systems/Top Events |
|---|---|
| Reactivity control and reactor sub-criticality | • Reactor trip (SCRAM and emergency boration) |
| Primary circuit integrity and pressure control | • Pilot operated relief valves (PORV)<br>• Feed and Bleed (F&B)<br>• Secondary side steam dump |
| Core re-flooding and primary inventory conservation | • High pressure injection system (HPIS)<br>• Re-flooding accumulators<br>• Low pressure injection system (LPIS) |

| Maintaining core cooling (including long-term residual heat removal) | <ul><li>Main Feedwater (MFW)</li><li>Emergency/auxiliary Feedwater (AFW)</li><li>Residual heat removal (RHR) system (both shutdown cooling and sump recirculation modes)</li><li>High pressure sump recirculation (HPR)</li></ul> |
|---|---|

Additionally considered events include:

- Emergency power supply failure (EPS).
- Early and late offsite power recovery.
- Service water system recovery.
- Ruptured steam generator isolation.
- Main steam-line isolation.
- Plant-level CCFs (as explained in section 2).
- Reactor coolant pump (RCP) seal LOCA.

Most of the modeled events -- especially for systems with multiple redundant trains -- have comprehensive fault trees quantifying their reliability following the modeling philosophy explained in section 2, covering frontline, support, human, and CCF contributions.

Table 2 shows a summary statistics of the structures used in our PWR PSA, having 77 fault trees, 11 event trees, and more than 2 million cutsets.

Table. 2. Summary statistics of the PWR PSA model

| Data Type | Number of Records |
|---|---|
| Fault Trees (Tops and Transfers) | 77 |
| Event Trees (Tops and Transfers) | 11 |
| Basic Events | 249 |
| Gates | 418 |
| Sequences | 114 |
| Cutsets | > 2,000,000 |

Figs. 1, 2, and 3 show some examples of our developed PWR event trees and fault trees. A full access to the PWR PSA models is available at Mendeley Data (http://dx.doi.org/10.17632/y9wfgyk6nz.1#folder-d062efaf-c881-46aa-b7c1-73be535bf5c0).

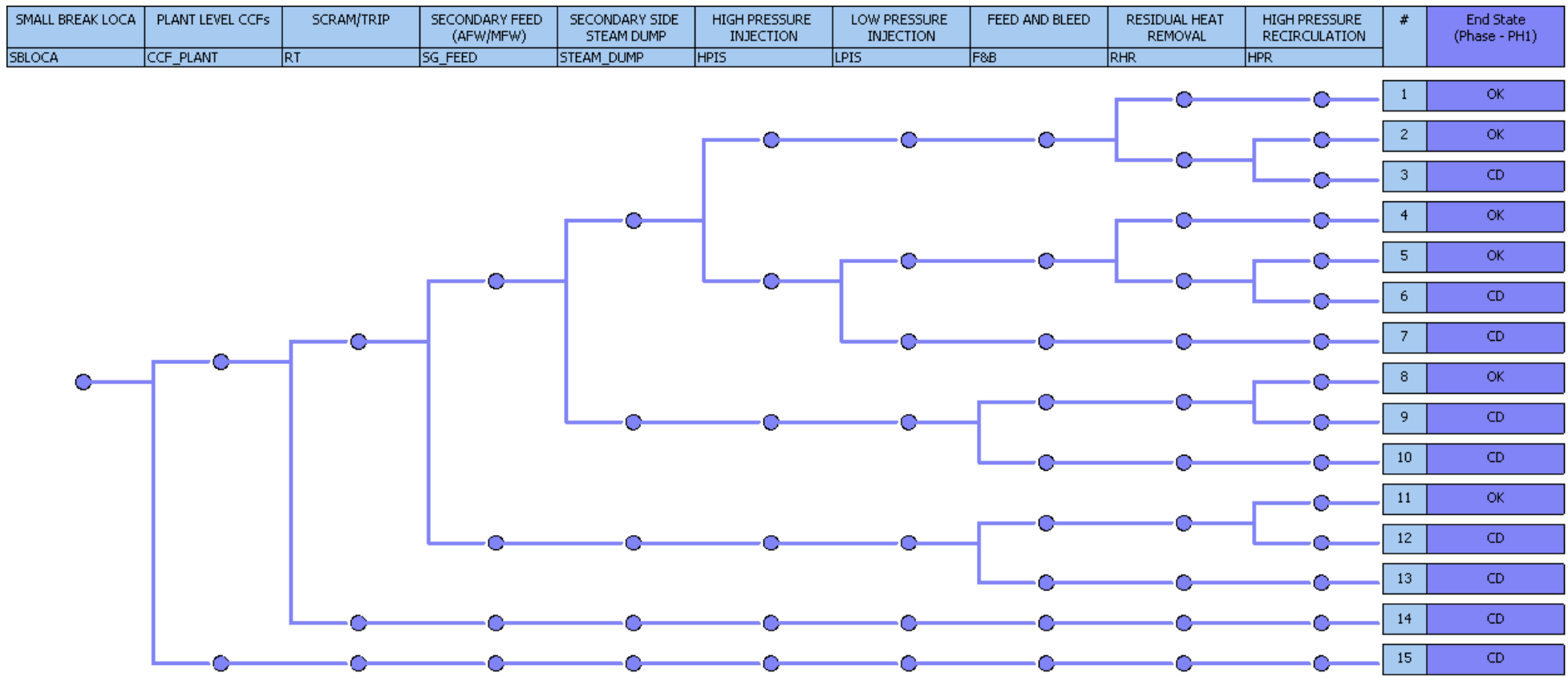

Fig. 1. An example of a developed PWR small-break LOCA event tree with two end-states: core damage (CD) and no core damage (OK)

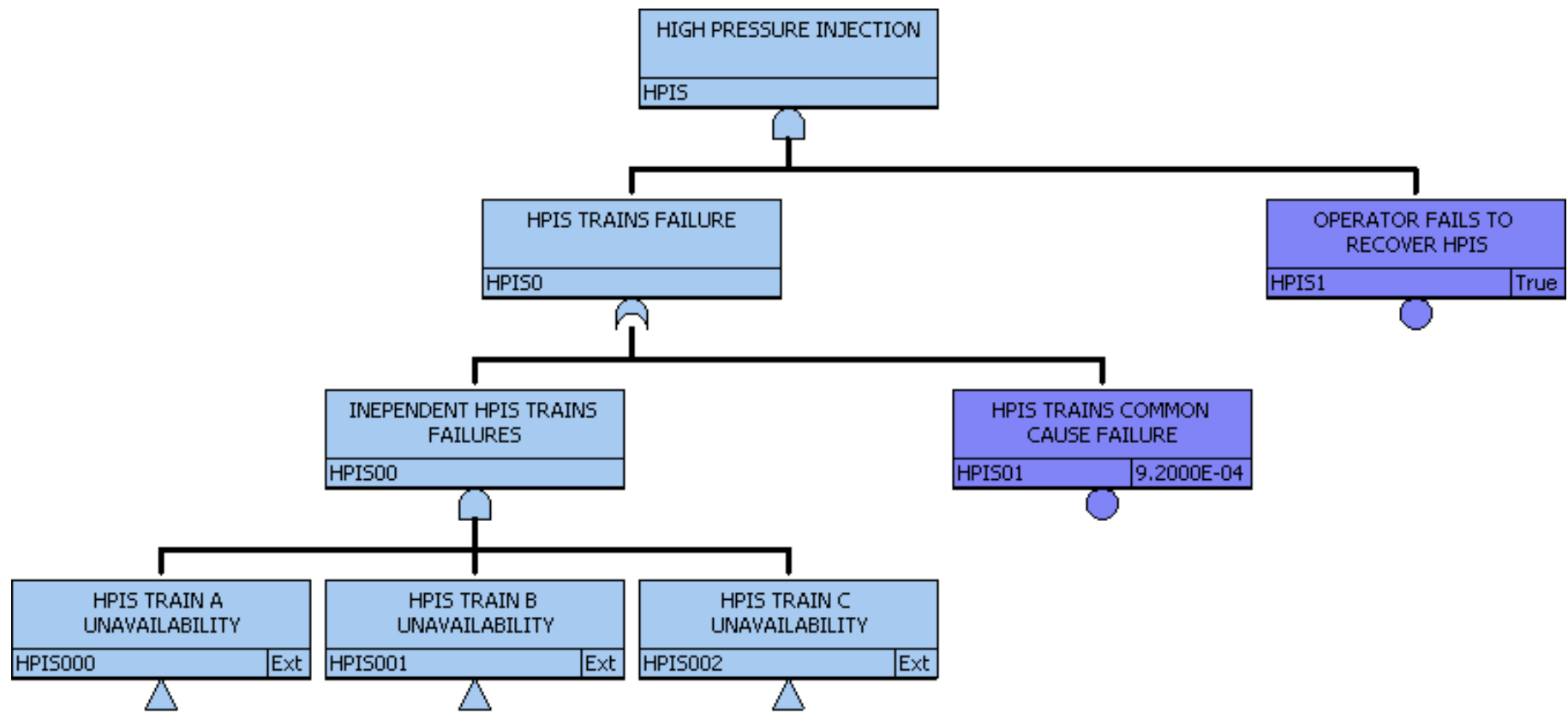

Fig. 2. An example of a developed PWR high-pressure injection system fault tree (see Fig. 3 for the transfer train fault tree)

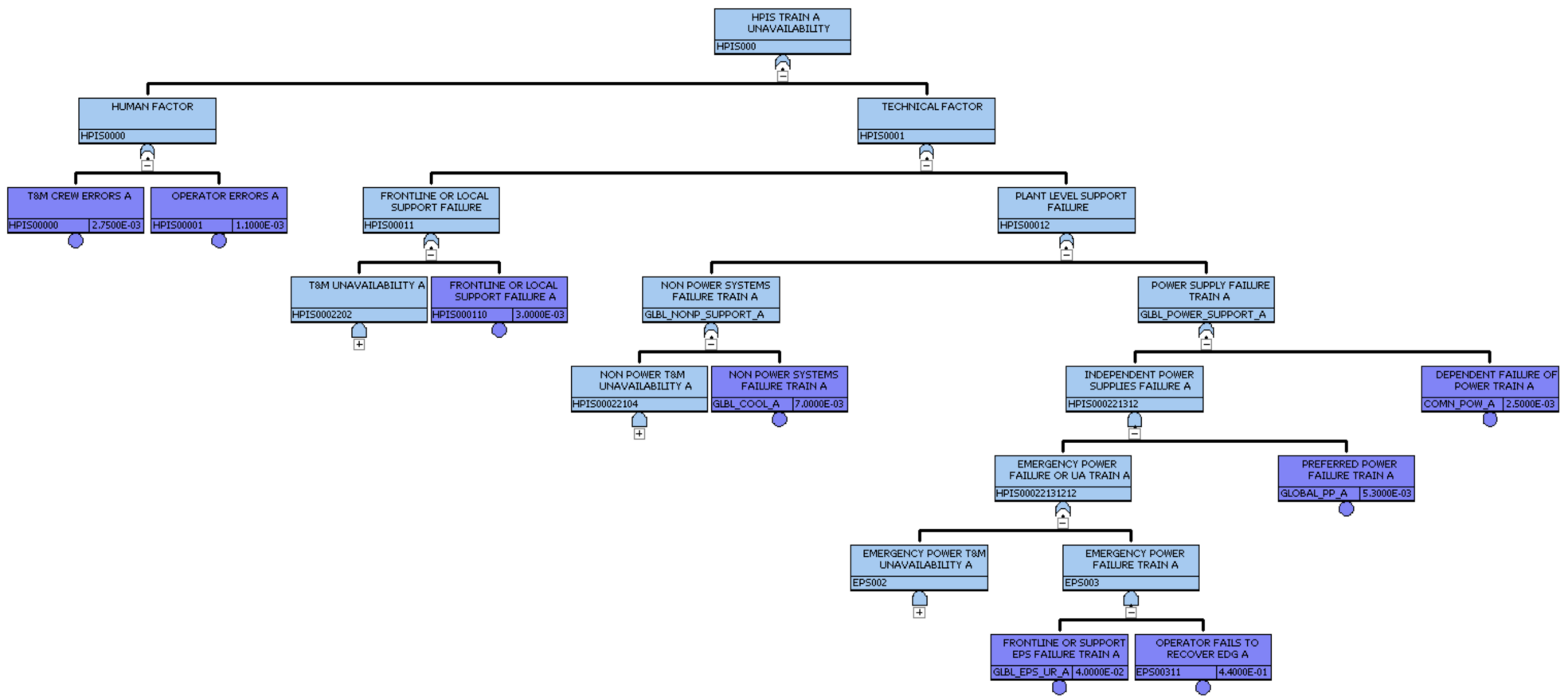

Fig. 3. Part of the PWR high-pressure injection train fault tree (for paper presentation purposes, additional branches are hidden under the '+' signs)

## 4. BWR models

For BWRs, we have considered internal initiating events similar to those modeled for PWRs with the exception of steam generator tube ruptures and main steam line breaks.

As in PWR models, an event tree is developed for each initiating event, encompassing safety systems/functions, and their backups. Table 3 lists these generic safety functions and the respective modeled safety systems/top events in our BWR PSA.

Table 3. Generic BWR safety functions and the associated modeled safety systems/top events

| Safety Functions | Safety Systems/Top events |
|---|---|
| Reactivity control and reactor sub-criticality | • Reactor trip (SCRAM and emergency boration) |
| Primary circuit integrity and pressure control | • Safety relief valves (SRVs) <br> • Reactor depressurization (automatic and manual) |
| Core re-flooding and primary inventory conservation | • Main Feedwater (MFW) <br> • Condensate system (COND) <br> • High pressure coolant injection (HPCI) <br> • High pressure core spray (HPCS) <br> • Reactor core isolation cooling (RCIC) <br> • Control rod drive injection (CRD) <br> • Low pressure coolant injection (LPCI) <br> • Low pressure core spray (LPCS) <br> • Alternative low pressure injection sources (fire, service water, or others) |
| Maintaining core cooling (including long-term residual heat removal) | • Isolation condenser (IC) <br> • Power conversion system (PCS) <br> • Residual heat removal (RHR) system (both shutdown cooling and suppression pool cooling modes) |

Additionally considered events include:

- Emergency power supply failure (EPS).
- Early and late offsite power recovery.
- Service water system recovery.
- Plant-level CCFs (as explained in section 2).

Most of the modeled events have comprehensive fault trees quantifying their reliability following the modeling philosophy explained in section 2, covering frontline, support, human, and CCF contributions.

Table 4 shows a summary statistics of the structures used in our BWR PSA, having 75 fault trees, 9 event trees, and more than 2 million cutsets.

Table. 4. Summary statistics of the BWR PSA model

| Data Type | Number of Records |
|---|---|
| Fault Trees (Tops and Transfers) | 75 |
| Event Trees (Tops and Transfers) | 9 |
| Basic Events | 224 |
| Gates | 221 |
| Sequences | 98 |
| Cutsets | > 2,000,000 |

Figs. A. 1 and A. 2 in the appendix show an example of a developed BWR event tree and fault tree. A full access to the BWR PSA models is available at Mendeley Data (http://dx.doi.org/10.17632/y9wfgyk6nz.1#folder-d062efaf-c881-46aa-b7c1-73be535bf5c0).

## 5. Data and quantification

For hardware unreliability data, and testing and maintenance unavailabilities (T&M UA), we are using US generic component reliability and availability data from the 2015 update of the NUREG/CR-6928 [21], aggregating components reliability parameters (both frontline and support components) to come up with a train failure probability estimate. The US, having the largest nuclear fleet, serves as a very good representative sample. An 8 hours mission time was used for the injection functions (e.g. HPIS), and 24 hours for the recirculation and decay heat removal phase (e.g. AFW, RHR) and global power and cooling support systems (e.g. EDGs, component-cooling pumps). Similarly, for initiating events frequencies and their uncertainty distributions, the 2015 update of the NUREG/CR-6928 was used.

For human error probabilities (EOO) and recovery actions, we used generic estimates similar to the classical methodologies of THERP, ATHENA, and the old USNRC precursor studies [22-25], depending on the feasibility of the operator action and recovery potentials (time, stress, prescriptive procedure, etc.). Regarding some basic events deduced from our database, such as EOC and T&M errors, we used their relative importance (occurrence frequency) in the database, as well as some expert's opinion, as proxies for their probabilities.

For computational purposes, we only model uncertainties in the most influential basic events based on their calculated importance measure (Fussell-Vesely, Risk reduction, Birnbaum Importance) [26], these include initiating events frequencies, plant-level CCFs, system-level CCFs, and EOOs. Furthermore, we also account for uncertainties in our database-estimated basic events (EOC, T&M errors, plant-level CCF).

Generally speaking, for system-level CCFs, a Beta-factor uniform distribution with [2%-20%] support (range) was adopted, including a worst-case scenario of 20% share of CCFs according to [27]. Going one level up, i.e. to plant-level CCF, the Beta-factor here is envisioned to go down 1 to 2 orders of magnitude (based on our database frequencies), therefore, we employ a plant-level Beta-factor uniform distribution with support [0.02%-2%] which is in line with the findings of [28], calculating a slightly less than 2% intersystem Beta-factor. Taking a classical system failure probability of about $10^{-4}$ per demand, a plant-level CCF uniform distribution with support [$2 \cdot 10^{-8}$ to $2 \cdot 10^{-6}$] is expected. For human errors and other uncertain parameters where no further information are given, a uniform (non-informative) distribution spanning one-order of magnitude is used following expert's opinion.

Table 5 presents all the modeled PWR and BWR initiating events along with their respective frequency distribution [21].

Table 5. PWR and BWR initiating events frequency distributions per reactor year (for explanation of acronyms, see section 3)

| Initiating Event | Frequency Distribution |
|---|---|
| General Transient (PWR) | Gamma ($\alpha$ = 7.9, $\beta$ = 11.6) |
| General Transient (PWR) | Gamma ($\alpha$ = 11.8, $\beta$ = 16) |
| LOCHS (PWR) | Gamma ($\alpha$ = 2.5, $\beta$ = 52) |
| LOCHS (BWR) | Gamma ($\alpha$ = 3.7, $\beta$ = 33) |
| SGTR (PWR) | Gamma ($\alpha$ = 2.5, $\beta$ = 1500) |
| MSLB (PWR) | Gamma ($\alpha$ = 10.5, $\beta$ = 1660) |
| SBLOCA PWR (sum of all small LOCAs) | Gamma ($\alpha$ = 0.4, $\beta$ = 535) |
| SBLOCA BWR (sum of all small LOCAs) | Gamma ($\alpha$ = 0.4, $\beta$ = 106) |
| MBLOCA (PWR) | Gamma ($\alpha$ = 0.3, $\beta$ = 1997) |
| MBLOCA (BWR) | Gamma ($\alpha$ = 0.4, $\beta$ = 4418) |
| LBLOCA (PWR) | Gamma ($\alpha$ = 0.3, $\beta$ = 50800) |
| LBLOCA (BWR) | Gamma ($\alpha$ = 0.3, $\beta$ = 25420) |
| LOOP | Gamma ($\alpha$ = 54.5, $\beta$ = 1750) |
| TLOSW | Gamma ($\alpha$ = 0.5, $\beta$ = 2032) |
| LONSW | Uniform ($5\ 10^{-4}$, $5\ 10^{-3}$)* |

*Experts opinion, $\alpha$ and $\beta$ are the Gamma distribution shape and rate parameters respectively

Table 6 shows some of the modeled basic events, that are common to both PWRs and BWRs, along with their respective nominal values/distributions. More examples are presented in the appendix (Tables A.1 and A.2). For brevity, we only present a sample of the whole list, and the reader is referred to the accompanying SAPHIRE PSA models for the full data.

Table 6. A sample of PWR and BWR common basic events and their nominal values

| Basic Event | Nominal Value or Distribution |
|---|---|
| General Train-level EOC | Uniform ($2\ 10^{-4}$, $2\ 10^{-3}$)** |
| General Train-level T&M Errors | Uniform ($5\ 10^{-4}$, $5\ 10^{-3}$)** |
| Operator Fails to Recover an EDG | Uniform ($8\ 10^{-2}$, $8\ 10^{-1}$)* |
| Early Offsite Power Recovery | Uniform ($5\ 10^{-2}$, $5\ 10^{-1}$)*[29] |
| Late Offsite Power Recovery | Uniform ($2\ 10^{-2}$, $1.2\ 10^{-1}$)*[29] |
| SW Recovery | Uniform ($1\ 10^{-3}$, $1\ 10^{-2}$)* |
| Operator Fails to Initiate Emergency Boration | Uniform ($5\ 10^{-3}$, $5\ 10^{-2}$)*[30] |
| Failure of a Safety Power Bus (Including Breaker Failure) | $2.5\ 10^{-3}$ |
| Global SW/CCW Train UR | $7\ 10^{-3}$ |
| Global SW/CCW Train UA | $1.3\ 10^{-2}$ |
| Operator Fails to Actuate RHR | Uniform ($5\ 10^{-4}$, $5\ 10^{-3}$) [31] |
| RHR CCF | Uniform ($1.7\ 10^{-4}$, $2\ 10^{-3}$) |

*Experts opinion; **Database-estimated basic events; CCW: Component Cooling Water

For a numerical exercise, we adapted our PWR PSA models to match a generic Westinghouse (W) PWR. Fig. 4 shows its CDF distribution generated using a large Latin hypercube sample of the model's uncertain parameters and basic events.

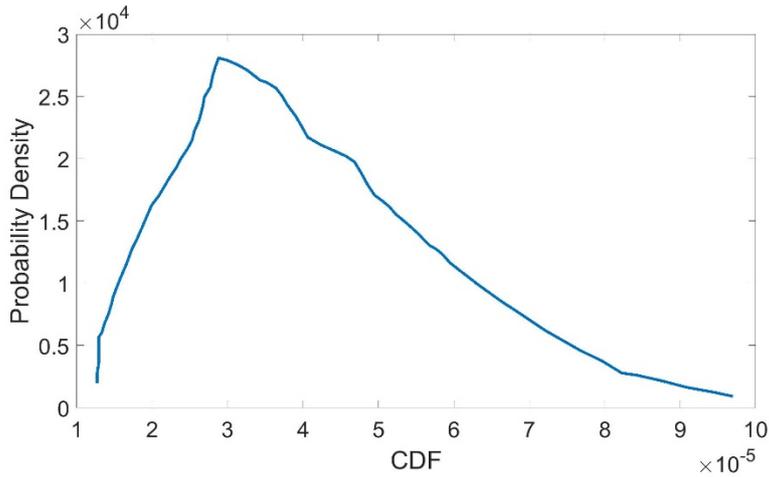

Fig. 4. Calculated CDF distribution of a generic PWR (W) with the following uncertainty statistics: mean $3.9 \cdot 10^{-5}$, standard deviation $1.6 \cdot 10^{-5}$, and 5th and 95th percentiles $1.7 \cdot 10^{-5}$ and $7 \cdot 10^{-5}$ respectively.

Table 7 shows the PSA results of different Westinghouse plants and a generic PWR found in the literature. The table presents the mean CDF, along with the 5th and 95th percentiles ($p_{05}$ and $p_{95}$) if available, thus serving as a benchmark for our generic PSA calculations.

Table 7. Literature PSA results of different PWRs (internal events only)

|  | **CDF $p_{05}$** | **Mean CDF** | **CDF $p_{95}$** |
|---|---|---|---|
| Surry (W) [32] | $6.8 \cdot 10^{-6}$ | $4 \cdot 10^{-5}$ | $1.3 \cdot 10^{-4}$ |
| Sequoyah (W) [32] | $1.2 \cdot 10^{-5}$ | $5.7 \cdot 10^{-5}$ | $1.8 \cdot 10^{-4}$ |
| Ringhals (W) [33] | NA | $2 \cdot 10^{-5}$ | NA |
| A Generic PWR [34] | NA | $3 \cdot 10^{-5}$ | NA |

For a typical General Electric (GE) BWR-4 plant design, our PSA model produces the following CDF distribution generated using Latin hypercube sampling (Fig. 5).

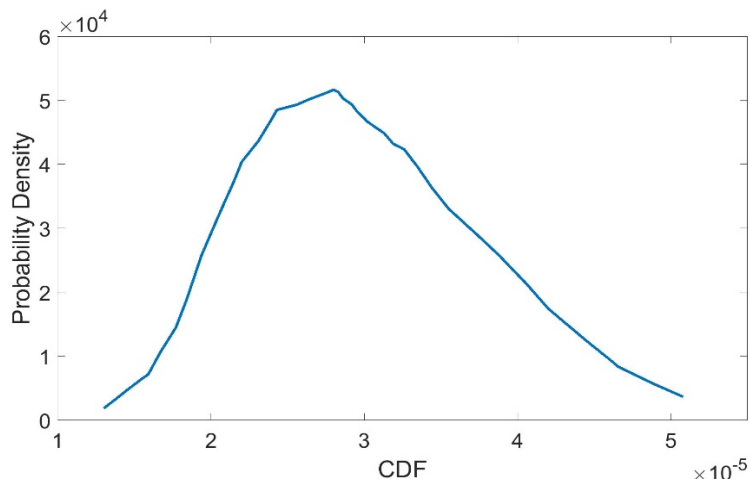

Fig. 5. Calculated CDF distribution of a typical GE BWR-4 design with the following uncertainty statistics: mean $2.9 \cdot 10^{-5}$, standard deviation $7.8 \cdot 10^{-6}$, and 5th and 95th percentiles $1.8 \cdot 10^{-5}$ and $4.3 \cdot 10^{-5}$ respectively.

Table 8 shows the PSA results of different BWR-4 plants and a generic BWR found in the literature. The table presents the mean CDF, along with the 5th and 95th percentiles ($p_{05}$ and $p_{95}$) if available, thus serving as a benchmark for our generic PSA calculations.

Table 8. Literature PSA results of different BWRs (internal events only)

|  | CDF $p_{05}$ | Mean CDF | CDF $p_{95}$ |
|---|---|---|---|
| Peach Bottom [33] | $2\ 10^{-7}$ | $4.5\ 10^{-6}$ | $9\ 10^{-5}$ |
| Forsmark [33] | NA | $1.1\ 10^{-5}$ | NA |
| A Generic BWR [35] | NA | $1.87\ 10^{-5}$ | NA |

Even though our models do not go to the plant-specific details, nor they are meant to replace industrial PSAs, the obtained results (Fig. 4 and 5) show a very good agreement with the literature ranges as depicted in Tables 7 and 8. In fact, the mean CDFs collected from the literature fit well within the calculated CDF distributions, and even more, our calculated CDF point estimates reasonably match the literature values shown in the tables.

Moreover, thanks to the easy arrangement and handiness of our PSA models, if the user is not satisfied with the generic estimates, and is interested in a more plant-specific experience, he/she can always zoom-in to the specific model parts, and adapt them or use more specific data, to get more satisfying results.

## 6. Precursor analysis and examples

As mentioned, one of the primary development goals of our PSA modeling approach is to have flexible models that are suitable to perform efficient and large-scale precursor analysis of events at different PWRs and BWRs around the world. In this section, we will give a hands-on precursor analysis illustration and application using our models.

To start with, a precursor analysis can be seen as a mapping of the empirical event or sequence of events at some plant to its respective PSA model. Therefore, if an event occurred, its corresponding basic event probability in the PSA model is adapted accordingly (either set to failure, i.e. probability of one, or modified depending on the degree of degradation). Now the question that a precursor analysis is interested in answering is: *what is the probability of arriving to a core damage due to the deterministic occurrence of that specific event?,* i.e. it is a conditional probability of core damage. This "conditional probability" serves as an estimate of the remaining defense against core damage after the observed events/failures have occurred, and hence it can be used to rank the risks of operational events.

Precursors can be of two types [36], the first involves a degraded plant condition (failure or degradation of systems/components) without the occurrence of an initiating event, and the other involves an initiating event with or without degraded plant conditions. For the first type, the probabilities of the basic events affected by the condition are modified accordingly, and hence, the adjusted PSA will be calculating a conditional core damage frequency (CCDF). A conditional core damage probability (CCDP) is then calculated as $CCDP = 1 - e^{-CCDF*T}$, where T is the condition duration. This CCDP is now compared with the base-case (i.e. unconditional) core damage probability $CDP_0$, defined as $CDP_0 = 1 - e^{-CDF_0*T}$, $CDF_0$ being the nominal core damage frequency. Finally, the risk metric for the degraded plant condition, for a duration $T$, is the incremental increase in core damage probability $\Delta CDP = CCDP - CDP_0$. For the second type of precursors, one of the postulated initiating events has actually happened, in addition, there might be some degraded plant conditions at the time of the initiating event. Therefore, for this type, the probability of the initiating event that has actually occurred is set to one, and all the other initiating events are set to zero. Additionally, the basic events affected by the condition are adjusted similar to the condition assessment case. The PSA model will now calculate a CCDP directly rather than a CCDF; this CCDP represents the risk metric for an initiating event precursor.

For events involving external initiators, the precursor analysis is done through capturing their manifested internal plant effects. For instance, if an earthquake led to some primary circuit pipe breaks (e.g. SBLOCA) and a failure of some equipment (e.g. EDGs), then the precursor risk is quantified by modelling a SBLOCA initiating event, and a concurrent EDG failure.

While performing the precursor analysis, plant design technicalities relevant to the event in hand are revisited and the respective models are adapted accordingly. Therefore, cross-tying, human intervention, and recovery possibilities will be credited whenever they are explicitly mentioned in the event under analysis. Furthermore, CCF basic event(s) of the affected system(s) are re-examined in case of CCF potential. Concretely, if, in a precursor, a component/train of a redundant system fails, and the remaining redundant components/trains had the potential to fail from the same cause, then $Q_t$ in the Beta-factor model[3] [37] is set to one, hence the CCF contribution $Q_n$ is $\beta . Q_t = \beta . 1 = \beta$.

To put some flesh on the concept, we hereby provide numerical examples of both precursor types. Note that our estimates tend to be fairly conservative as we generally do not account for systems recovery in the nominal case (as previously discussed).

**Degraded plant condition precursor:**

Take a precursor at a (W) PWR involving a non-recoverable one-month unavailability of a turbine driven auxiliary feedwater (TDAFW) pump. Assuming the AFW system of this plant consists of 1 turbine driven pump and 2 motor driven pumps, our PSA model of this precursor generates the CCDF uncertainty distribution shown in Fig. 6, with mean of 7.4 10$^{-5}$/year.

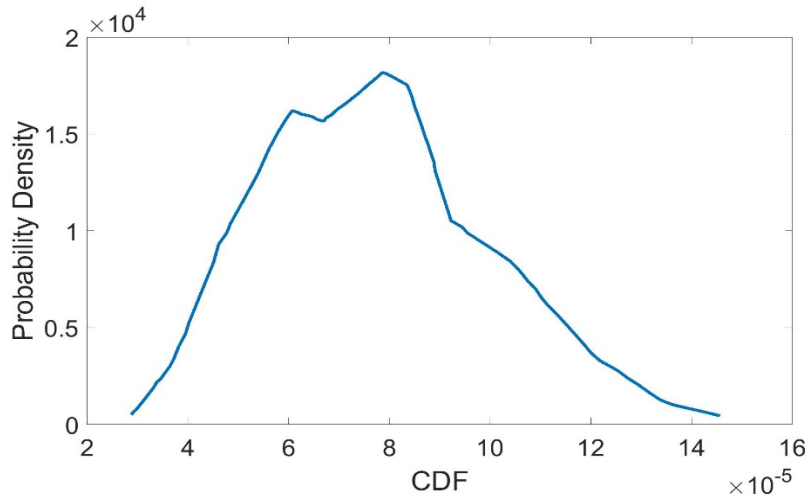

Fig. 6. CCDF distribution of a PWR (W) TDAFW pump unavailability event

$$CCDP \approx CCDF * T = \left(\frac{7.4\text{E}^{-5}}{\text{year}}\right) \frac{1\ month}{12\ \frac{months}{year}} = 6.1\ 10^{-6}.$$

The $CDF_0$ for a generic (W) PWR is 3.9 10$^{-5}$ (Fig. 6), hence, $CDP_0 \approx CDF_0 * T = 3.2\ 10^{-6}$. Therefore, the incremental increase in core damage probability during the one-month exposure to this precursor is $\Delta CDP = 2.9\ 10^{-6}$, consequently, an incremental increase in large early release probability $\Delta LERP \sim 2.9\ 10^{-7}$ is expected.

---

[3] $\beta = \frac{Q_n}{Q_1 + Q_n} = \frac{Q_n}{Q_t}$, where $Q_n$ is the CCF contribution (failure probability due to totally dependent failures), $Q_1$ is the component (i.e. train) failure probability due to independent causes, $Q_t$ is the total failure probability of a component (i.e. train) both from independent and dependent causes.

**Initiating event precursor:**

Take a precursor at a GE BWR involving a LOOP and a failure of 1 out of 2 available EDGs – with recovery. Our PSA model of this precursor – for a generic BWR/4/5/6 production series – calculates the CCDP uncertainty distribution shown in Fig. 7, with mean CCDP of $1.3 \cdot 10^{-3}$ (hence a conditional LERP ~ $1 \cdot 10^{-4}$).

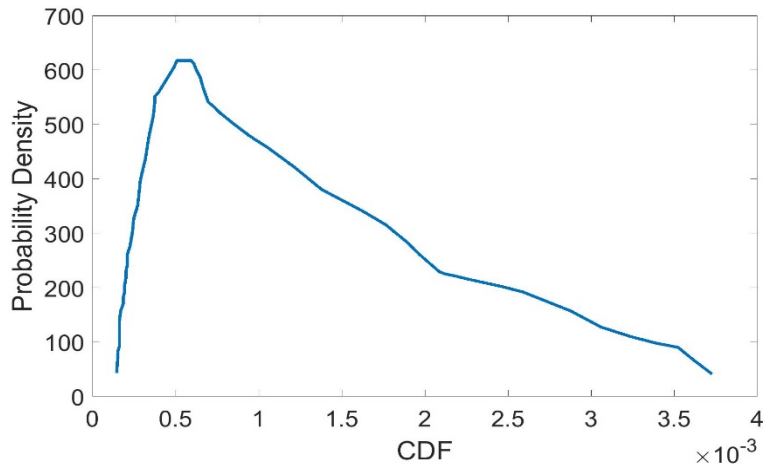

Fig. 7. CCDP distribution of a GE BWR LOOP event concurrent with an EDG unavailability

## 7. Conclusions and future work

In this work, we have presented the final results of our in-house generic PSA models for PWRs and BWRs. These models cover all common internal initiating events, with intermediate complexity generic event trees and fault trees capturing major design differences (redundancy, automation level, support dependence, etc.) without going to the details. This mentality of going simple and generic allowed for accounting for important lessons learnt from hundreds of precursors in our curated ETHZ Curated Nuclear Events Database [18], hence, supporting the quest for PSA completeness.

The developed models have attractive characteristics and are adaptable to account for plant-specific differences and new factors (EOC, T&M errors, Plant CCF, and others). Thanks to these characteristics, the models are foreseen to:

- Serve as a complementary framework that can aid plant-specific PSAs and answer big picture questions safety concerns.
- Provide an efficient platform for order-of-magnitude precursor analysis.
- Act as a fast first filter for precursors that can be used for initial screening and generic risk insights.
- Offer an unbiased framework – by annealing-out many plant-specific differences (as a result of going generic) – to compare precursors and risks at different plants, hence, help to learn and suggest cost-effective back-fits.
- Provide PSA models that could be used for different applications, when no access to plant-specific PSAs is provided (research institutes, universities, public, etc…).
- Help new comers and developing countries, serving as a starting point for their plant-specific PSAs as they are easy to understand and adapt.
- Support the PSA harmonization goal.

The models are now deployed for a large-scale precursor analysis for about 1000 worldwide safety-relevant events from our database. The unique outcome of this effort will be used to generate many empirical and statistical lessons, assess nuclear operational risks, and provide a framework for precursor's simulation and accident's prediction. It will help to observe developing minor failures over time to identify and monitor signals that may announce catastrophic collapse, and eventually develop active control and possible remedies through risk-informed decision-making.

**Acknowledgement**

The authors would like to acknowledge the continuous support provided by the Gösgen and the Leibstadt nuclear power plants in Switzerland throughout the project.

**Appendix:**

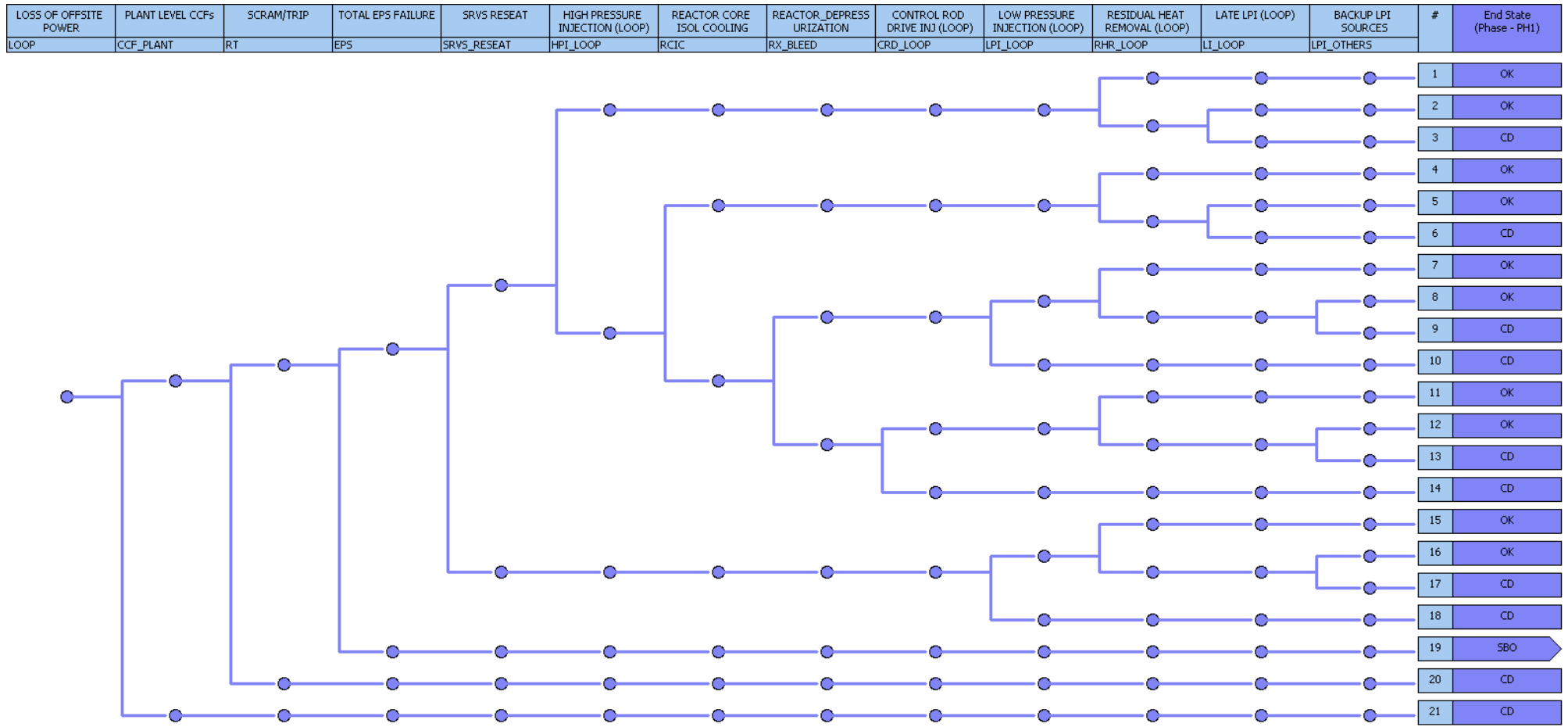

Fig. A.1. An example of a developed BWR LOOP event tree with two end-states: core damage (CD) and no core damage (OK)

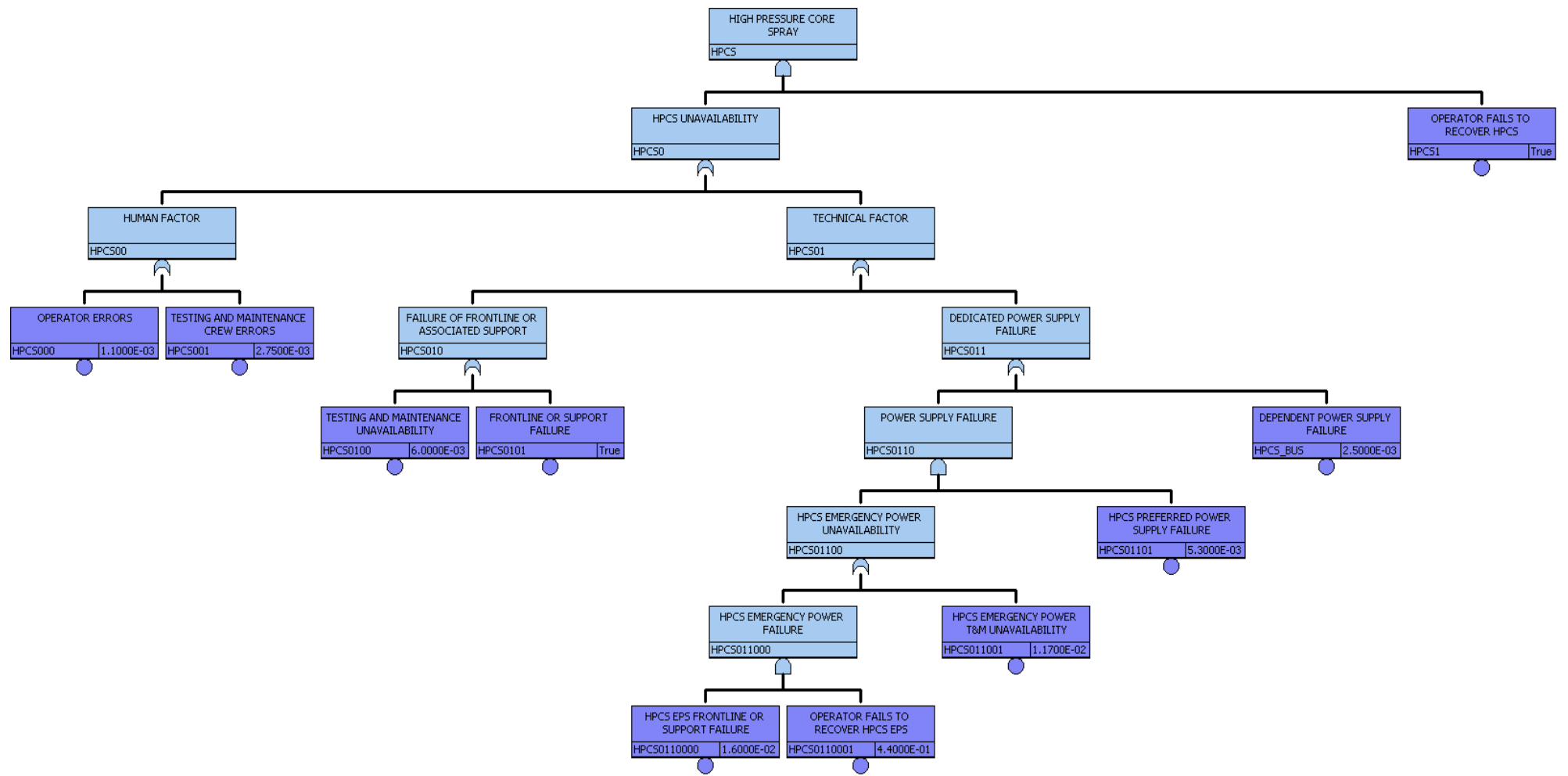

Fig. A.2. An example of a developed BWR high-pressure core spray fault tree

Table A.1. A sample of PWR basic events and their nominal values

| Basic Event | Nominal Value or Distribution |
|---|---|
| Operator Fails to Trip the RCP Following Loss of Seal Cooling | Uniform ($2\ 10^{-2}$, $2\ 10^{-1}$) [30] |
| PORV Reseat Recovery Failure – Including Block Valves | Uniform ($5\ 10^{-3}$, $5\ 10^{-2}$) [30] |
| AFW CCF | Uniform ($1.8\ 10^{-4}$, $2.2\ 10^{-3}$) |
| Operator Fails to Actuate/Reconnect MFW | Uniform ($5\ 10^{-4}$, $5\ 10^{-3}$)*[30] |
| Operators Fail to Open Steam Dump Valves | Uniform ($2\ 10^{-4}$, $2\ 10^{-3}$)*[31] |
| HPIS CCF | Uniform ($1.4\ 10^{-4}$, $1.7\ 10^{-3}$) |
| LPIS CCF | Uniform ($1.4\ 10^{-4}$, $1.7\ 10^{-3}$) |
| Operator Fails to Actuate HPIS in F&B | Uniform ($2\ 10^{-3}$, $2\ 10^{-2}$) [30] |
| Operators Fails to Detect and Isolate Broken Steam Line | Uniform ($1\ 10^{-3}$, $1\ 10^{-2}$)* |
| Operators Fails to Detect and Isolate Ruptured Steam Generator | Uniform ($1\ 10^{-3}$, $1\ 10^{-2}$)* |

*Experts opinion

Table A.2. A sample of BWR basic events and their nominal values

| Basic Event | Nominal Value or Distribution |
|---|---|
| SRVs Fail to Reset | $1\ 10^{-3}$* |
| HPCI Frontline and Local Support Single Train UR | $5.1\ 10^{-2}$ [21] |
| RCIC Frontline and Local Support Single Train T&M UA | $1.1\ 10^{-2}$ [21] |
| ADS Failure | $3.7\ 10^{-3}$ [30] |
| Operator Fails to Manually Depressurize the Reactor | $7\ 10^{-1}$ [30] |
| Operator Fails to Align CRD | Uniform ($5\ 10^{-3}$, $5\ 10^{-2}$) [31] |
| LPCI CCF | Uniform ($1.4\ 10^{-4}$, $1.7\ 10^{-3}$) |
| Operators Fail to Align Alternative Low Pressure Injection Sources | Uniform ($3\ 10^{-3}$, $3\ 10^{-2}$)*[31] |

*Experts opinion; ADS: Automatic Depressurization System